\pdfoutput=1
\documentclass[12pt]{article}
\usepackage{authblk}
\usepackage[bookmarksnumbered, colorlinks, plainpages]{hyperref}
\usepackage{hyperref}
\usepackage{amsmath, amsthm, amscd, amsfonts, amssymb, graphicx, color, booktabs}
\usepackage[a4paper, total={6.5in, 10.4in}]{geometry}
\usepackage{xspace} 
\usepackage{upgreek}

\def\lhcb   {\mbox{LHCb}\xspace}

\def\MagUp {\mbox{\em Mag\kern -0.05em Up}\xspace}

{

 \def\PDelta      {\ensuremath{\Delta}\xspace}                 
 \def\PXi         {\ensuremath{\Xi}\xspace}                 
 \def\PLambda     {\ensuremath{\Lambda}\xspace}                 
 \def\PSigma      {\ensuremath{\Sigma}\xspace}                 
 \def\POmega      {\ensuremath{\Omega}\xspace}                 
 \def\PUpsilon    {\ensuremath{\Upsilon}\xspace}
 \let\oldPi\Pi
 \def\PPi         {\ensuremath{\oldPi}\xspace}

 \def\PB      {\ensuremath{\mathrm{B}}\xspace}                 
                  
 \def\PD      {\ensuremath{\mathrm{D}}\xspace}

 \def\PK      {\ensuremath{\mathrm{K}}\xspace}

 \def\Pi      {\ensuremath{\mathrm{i}}\xspace}

 \def\Ps      {\ensuremath{\mathrm{s}}\xspace}

 \def\thebaroffset{0.0em}
}
{

 \mathchardef\PDelta="7101
 \mathchardef\PXi="7104
 \mathchardef\PLambda="7103
 \mathchardef\PSigma="7106
 \mathchardef\POmega="710A
 \mathchardef\PUpsilon="7107
 \mathchardef\PPi="7105
                  
 \def\PB      {\ensuremath{B}\xspace}                 
                  
 \def\PD      {\ensuremath{D}\xspace}

 \def\PK      {\ensuremath{K}\xspace}

 \def\Pi      {\ensuremath{i}\xspace}

 \def\Ps      {\ensuremath{s}\xspace}

 \def\thebaroffset{0.18em}
}
\newcommand{\offsetoverline}[2][\thebaroffset]{\kern #1\overline{\kern -#1 #2}}%

\makeatletter
\ifcase \@ptsize \relax
  \newcommand{\miniscule}{\@setfontsize\miniscule{4}{5}}
\or
  \newcommand{\miniscule}{\@setfontsize\miniscule{5}{6}}
\or
  \newcommand{\miniscule}{\@setfontsize\miniscule{5}{6}}
\fi
\makeatother

\DeclareRobustCommand{\optbar}[1]{\shortstack{{\miniscule (\rule[.5ex]{1.25em}{.18mm})}
  \\ [-.7ex] $#1$}}

\def\squark    {{\ensuremath{\Ps}}\xspace}

\def\KorKbar {\kern \thebaroffset\optbar{\kern -\thebaroffset \PK}{}\xspace}

\def\D       {{\ensuremath{\PD}}\xspace}

\def\DorDbar {\kern \thebaroffset\optbar{\kern -\thebaroffset \PD}\xspace}

\def\Dp      {{\ensuremath{\D^+}}\xspace}
\def\Dm      {{\ensuremath{\D^-}}\xspace}

\def\DpDm    {\ensuremath{\Dp {\kern -0.16em \Dm}}\xspace}

\def\B       {{\ensuremath{\PB}}\xspace}

\def\BorBbar {\kern \thebaroffset\optbar{\kern -\thebaroffset \PB}\xspace}

\def\Bd      {{\ensuremath{\B^0}}\xspace}

\def\BdorBdbar {\kern \thebaroffset\optbar{\kern -\thebaroffset \Bd}\xspace}

\def\Bs      {{\ensuremath{\B^0_\squark}}\xspace}

\def\BsorBsbar {\kern \thebaroffset\optbar{\kern -\thebaroffset \Bs}\xspace}

\def\Y#1S{\ensuremath{\PUpsilon{(#1S)}}\xspace}

\def\LorLbar     {\kern \thebaroffset\optbar{\kern -\thebaroffset \PLambda}\xspace}

\def\to                 {\ensuremath{\rightarrow}\xspace}

\def\CP                {{\ensuremath{C\!P}}\xspace}

\def\AT#1     {\ensuremath{A_{\mathrm{T}}^{#1}}\xspace}           

\def\C#1      {\ensuremath{\mathcal{C}_{#1}}\xspace}                       
\def\Cp#1     {\ensuremath{\mathcal{C}_{#1}^{'}}\xspace}                    
\def\Ceff#1   {\ensuremath{\mathcal{C}_{#1}^{\mathrm{(eff)}}}\xspace}        
\def\Cpeff#1  {\ensuremath{\mathcal{C}_{#1}^{'\mathrm{(eff)}}}\xspace}       
\def\Ope#1    {\ensuremath{\mathcal{O}_{#1}}\xspace}                       
\def\Opep#1   {\ensuremath{\mathcal{O}_{#1}^{'}}\xspace}                    

\newcommand{\aunit}[1]{\ensuremath{\text{\,#1}}}       

\newcommand{\tev}{\aunit{Te\kern -0.1em V}\xspace}
\newcommand{\gev}{\aunit{Ge\kern -0.1em V}\xspace}
\newcommand{\mev}{\aunit{Me\kern -0.1em V}\xspace}
\newcommand{\kev}{\aunit{ke\kern -0.1em V}\xspace}
\newcommand{\ev}{\aunit{e\kern -0.1em V}\xspace}
 
\newcommand{\mevc}{\ensuremath{\aunit{Me\kern -0.1em V\!/}c}\xspace}
\newcommand{\gevc}{\ensuremath{\aunit{Ge\kern -0.1em V\!/}c}\xspace}
\newcommand{\mevcc}{\ensuremath{\aunit{Me\kern -0.1em V\!/}c^2}\xspace}
\newcommand{\gevcc}{\ensuremath{\aunit{Ge\kern -0.1em V\!/}c^2}\xspace}

\def\gsim{{~\raise.15em\hbox{$>$}\kern-.85em
          \lower.35em\hbox{$\sim$}~}\xspace}
\def\lsim{{~\raise.15em\hbox{$<$}\kern-.85em
          \lower.35em\hbox{$\sim$}~}\xspace}

\def\degrees{\ensuremath{^{\circ}}\xspace}

\def\tell1  {TELL1\xspace}
\def\ukl1   {UKL1\xspace}

\newcommand{\lhcborcid}[1]{\href{https://orcid.org/#1}{\hspace*{0.1em}\raisebox{-0.45ex}{\includegraphics[width=1em]{figs/orcidIcon.pdf}}}}

\usepackage{bibspacing}
\numberwithin{equation}{section}
\definecolor{email}{rgb}{0.00,0.00,0.84}
\begin{document}
\setcounter{page}{1}
\renewcommand{\theequation}{\arabic{equation}}

\title{\vspace{-1cm}\large \bf 12th Workshop on the CKM Unitarity Triangle\\ Santiago de Compostela, 18-22 September 2023 \\ \vspace{0.3cm}
\Large New results of $\gamma$ measurements in ADS and GLW-like decays at LHCb}

\author[1]{\normalsize Seophine Stanislaus on behalf of the LHCb collaboration\thanks{seophine.stanislaus@cern.ch}}
\affil[1]{\normalsize Department of Physics, University of Oxford, Oxford, United Kingdom}
\date{March 28, 2024}

\maketitle
\vspace{-1cm}
\begin{abstract}
Recent LHCb measurements of the CKM angle $\gamma$ in ADS and GLW-like decays are presented. One measurement considers $B^{0}\to D K^{*0}(892)$ decays with two- and four-body $D$-decay final states and another considers $B^{\pm}\to D h^{\pm}$ decays with $D \to K\pi\pi\pi$ final states. The former supersedes a previous LHCb measurement and was presented for the first time at the 12th Workshop on the CKM Unitarity Triangle. The latter analysis measures the largest \ensuremath{C\!P} asymmetry seen at LHCb by splitting the $D$-decay phase space into bins. 
\end{abstract} \maketitle

\section*{Introduction}

\noindent The Standard Model describes \CP-violation in the quark sector using the Cabibbo-Kobayashi-Maskawa (CKM) matrix~\cite{Cabibbo:1963yz, Kobayashi:1973fv} which can be represented using the Unitarity Triangle~\cite{wolfenstein1983parametrization}. In this triangle, measurements of the CKM angle $\gamma$ are particularly important as it is the only angle that can be measured using only tree-level processes which are assumed not to include physics beyond the Standard Model~\cite{Brod_2015,Lenz:2019lvd}. Therefore, direct measurements of $\gamma$ can be compared to indirect determinations which are sensitive to loop contributions from new physics to test the Standard Model. Measurements of CKM angle $\gamma$ are also useful as they have negligible theoretical uncertainty since all hadronic parameters are determined from data~\cite{Brod_2014}. 

The \lhcb collaboration has performed numerous measurements of $\gamma$ where direct \CP-violation is measured using the interference of $b \to u \bar{c} s$ and $b \to c \bar{u} s$ quark transitions in, for example, ${B^{\pm} \to D K^{\pm}}$ decays. The weak interaction vertices involved allow access to $\gamma$ which can be approximated to be $\arg\left[V_{cb}V_{us}^{*}/V_{ub}V_{cs}^{*}\right]$~\cite{Grossman_2014}. The \lhcb detector~\cite{LHCb-DP-2008-001,LHCb-DP-2014-002} is well-suited for this topology of decay. The Vertex Locator (VELO) provides excellent vertex resolution to identify displaced $B$ and $D$ vertices. Particle identification is also a vital component and is achieved using the VELO, tracking system and, most importantly, the RICH detectors. 

Having studied numerous charged and neutral $B$ decays, the latest combination of $\gamma$ measurements from \lhcb produced a value of $\gamma=(63.8^{+3.5}_{-3.7})\degrees$~\cite{LHCb-CONF-2022-003}. To fully exploit a $B$ decay, it is important to consider different $D$ final states since each gives different sensitivity to different hadronic parameters. This work presents two recent results from \lhcb~\cite{Gilman:2812327,2023} which studied two- and four-body $D$ final states and \CP-violation is measured using the ADS~\cite{PhysRevLett.78.3257,PhysRevD.63.036005} and GLW~\cite{GRONAU1991172,GRONAU1991483} methods.

\section*{$B^{0}\to D K^{*0}(892)$ Decays}

A measurement of \CP-violation in $B^{0}\to D K^{*0}(892)$ decays with two- and four-body $D$ final states was recently completed by \lhcb~\cite{Gilman:2812327}. The measurement supersedes results from a previous \lhcb Run 1 and partial Run 2 analysis~\cite{LHCb-PAPER-2019-021}, which corresponded to a total integrated luminosity of 4.8 fb$^{-1}$. The measurement presented here uses the full 9 fb$^{-1}$ Run 1 and Run 2 dataset.

When considering $B^{0}\to D K^{*0}(892)$ decays, the charge on the kaon from the $K^{*0}$ meson is used to tag the flavour of the $B$ meson at decay. This results in a time-integrated measurement with no effects from $B^{0}$--$\overline{B^{0}}$ mixing. In the case of $B^{0}\to D K^{*0}(892)$ decays, although the $K^{*0}$ resonance is selected, there are also non-resonant contributions. This means that a coherence factor, $\kappa$, is necessary to account for the small contribution from other intermediate states. The value for this is taken from Ref.~\cite{LHCb-PAPER-2015-059}, and therefore the same selection for the $K^{*0}$ meson is used. Details on the other selections can be found in Ref.~\cite{Gilman:2812327}.

\subsection*{GLW-like decays}
\begin{figure}
    \centering
    \includegraphics[width=0.43\textwidth]{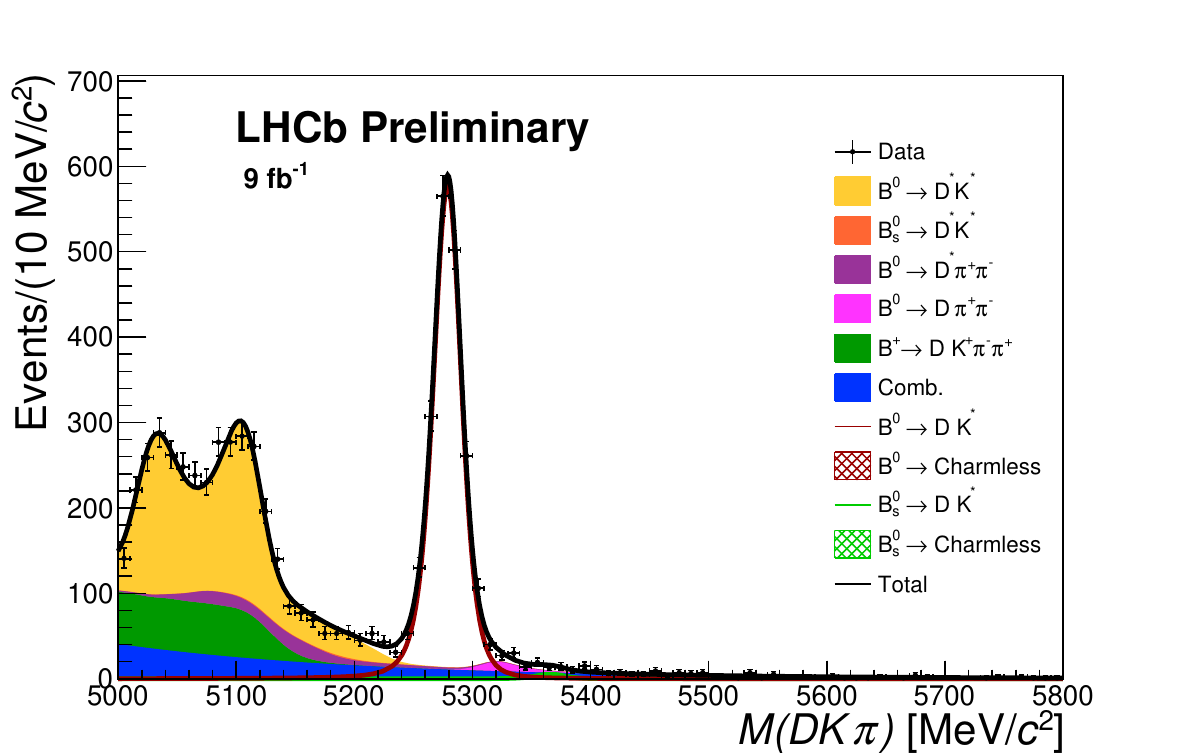}\hfill
    \includegraphics[width=0.43\textwidth]{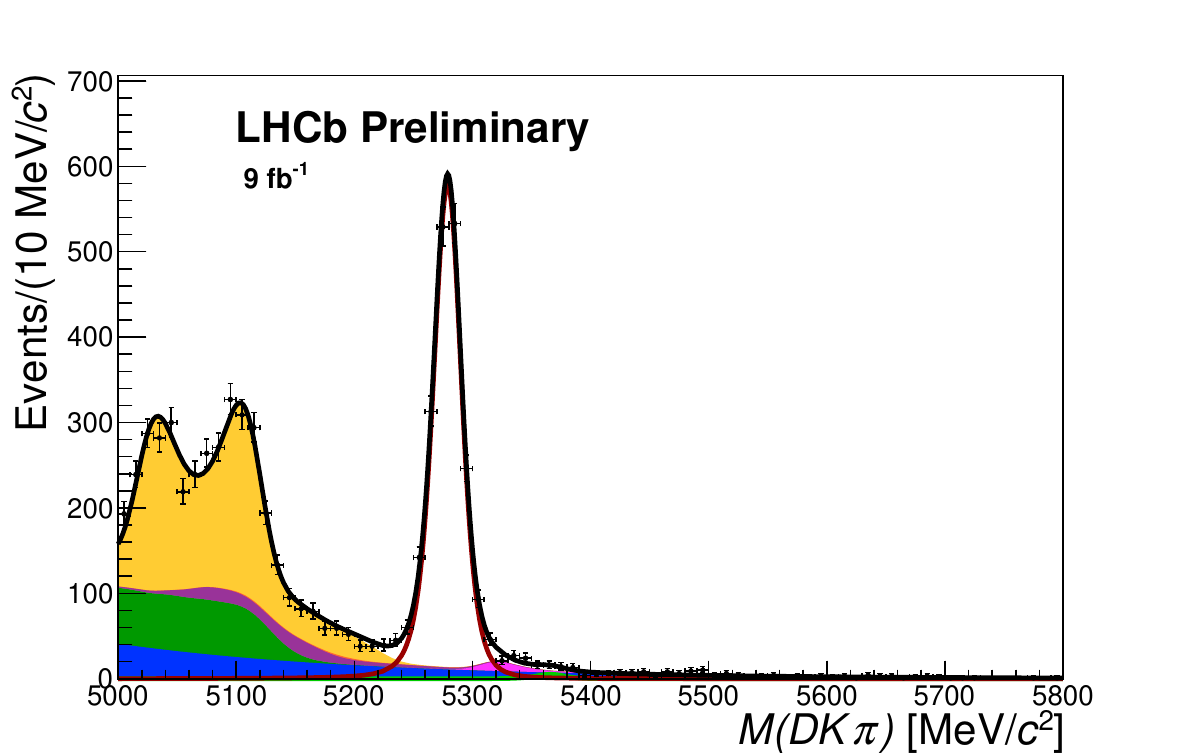}\\
    \includegraphics[width=0.43\textwidth]{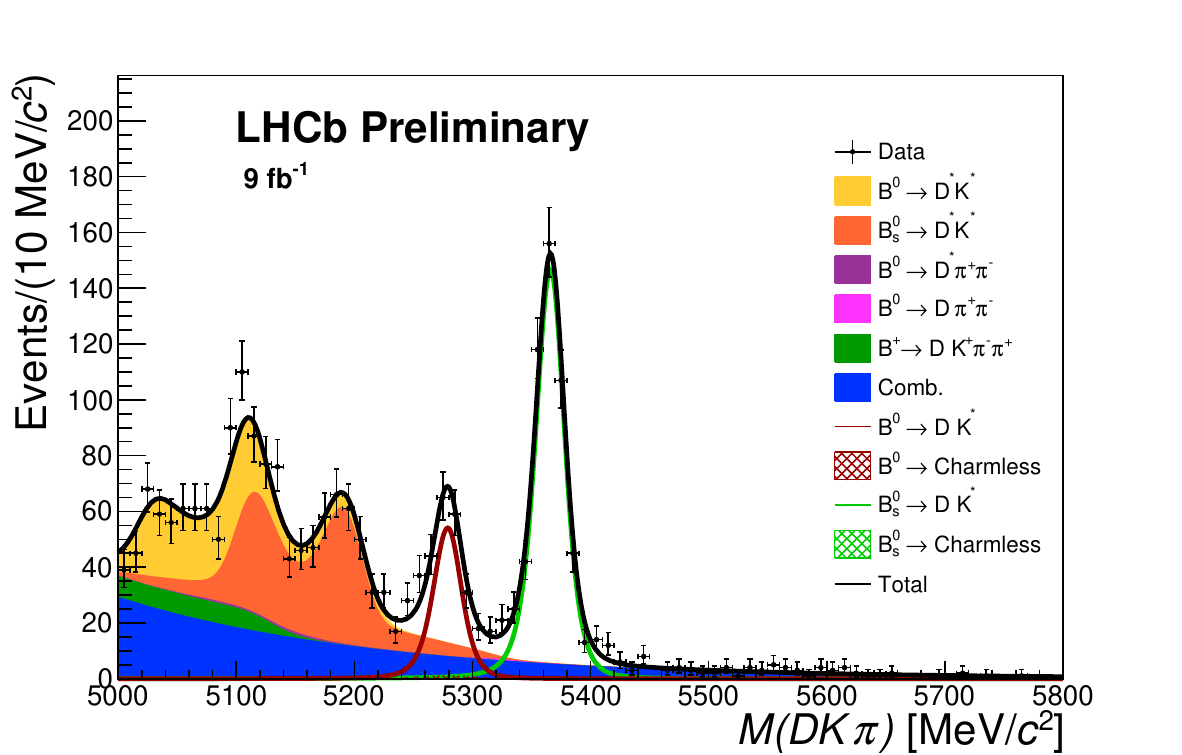}\hfill
    \includegraphics[width=0.43\textwidth]{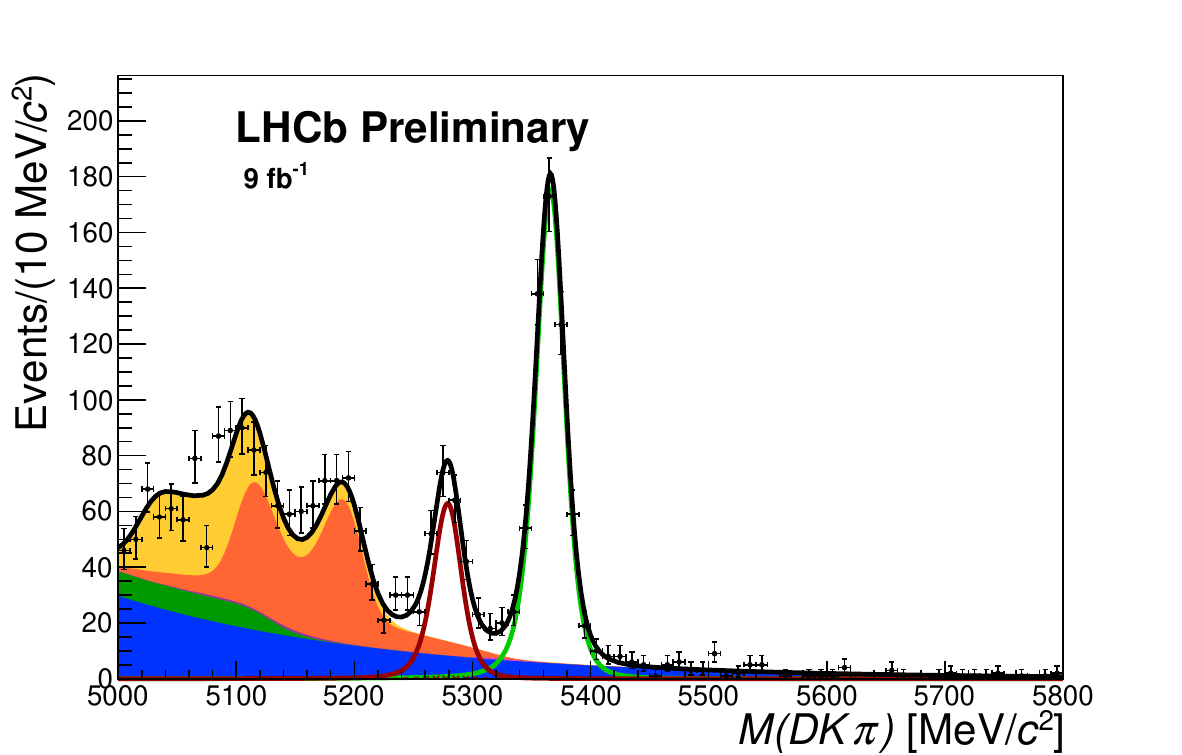} \\
    \includegraphics[width=0.43\textwidth]{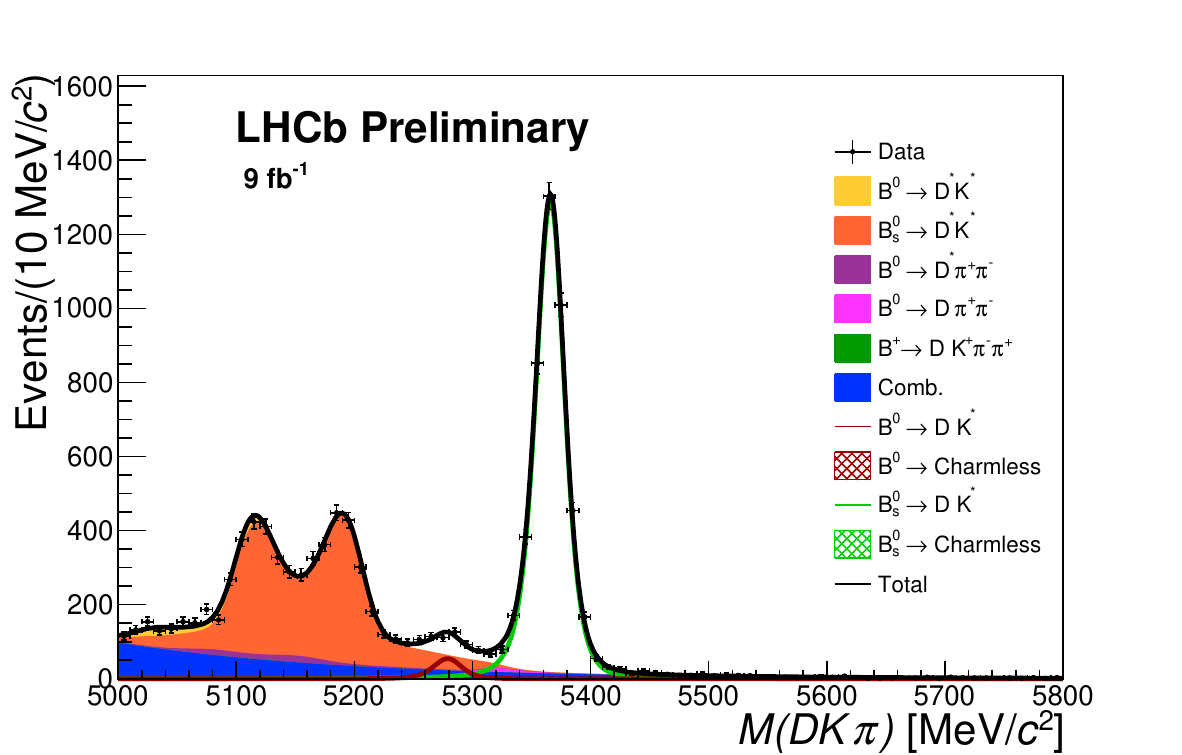}\hfill
    \includegraphics[width=0.43\textwidth]{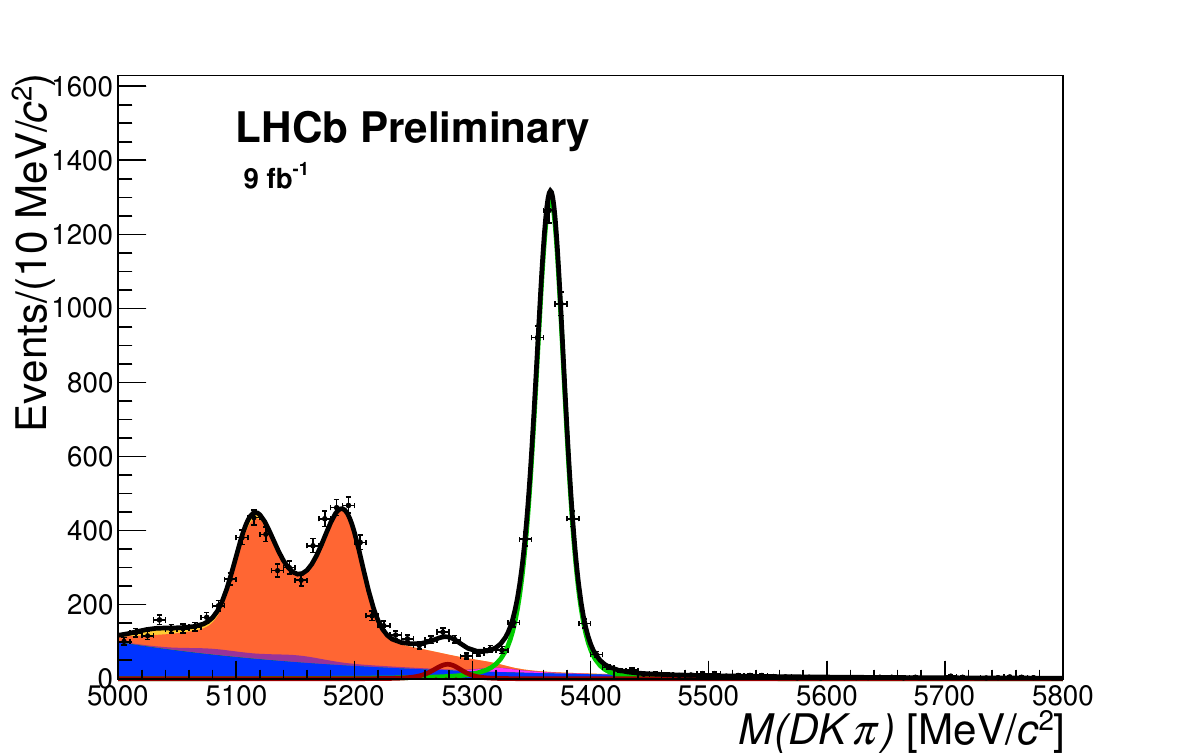} \\
        \includegraphics[width=0.43\textwidth]{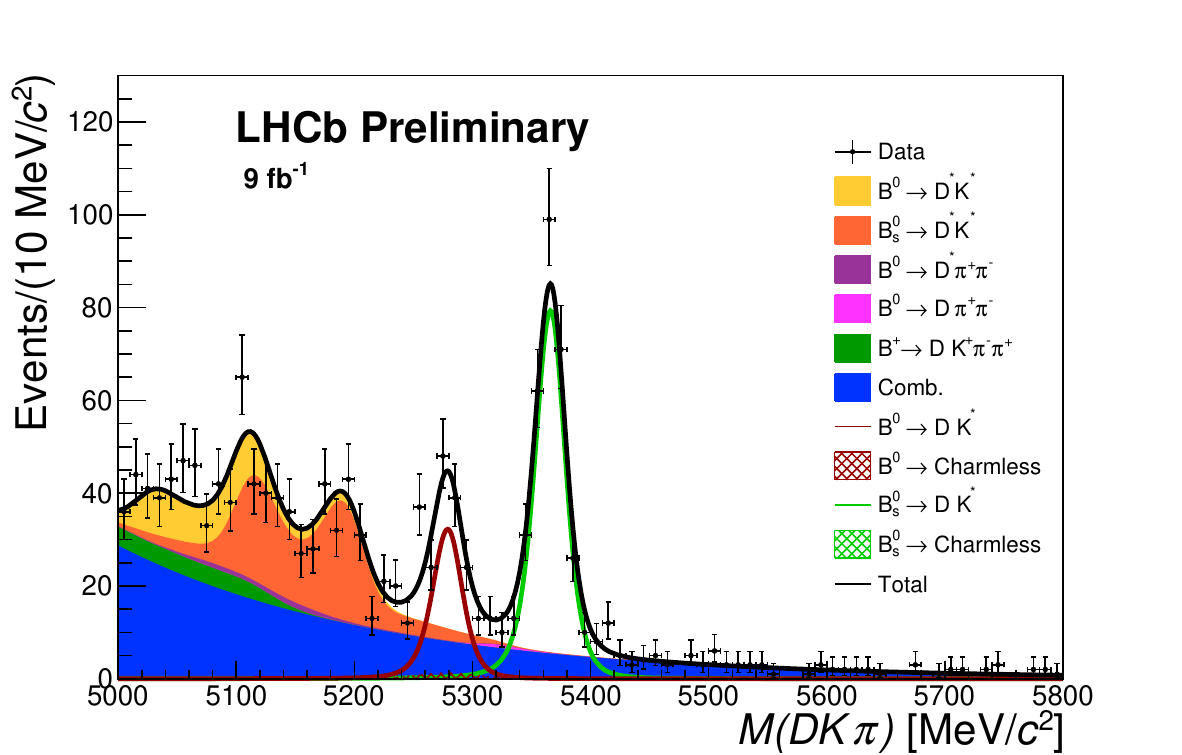}\hfill
    \includegraphics[width=0.43\textwidth]{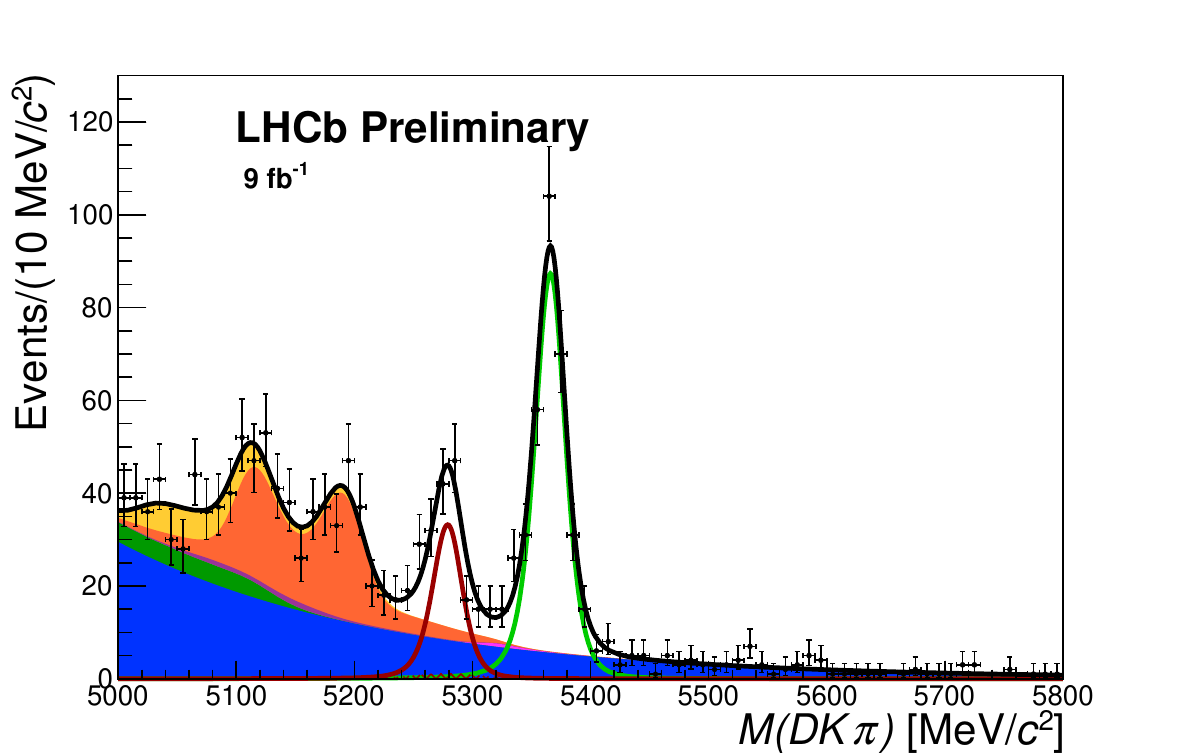}    \\
    \includegraphics[width=0.43\textwidth]{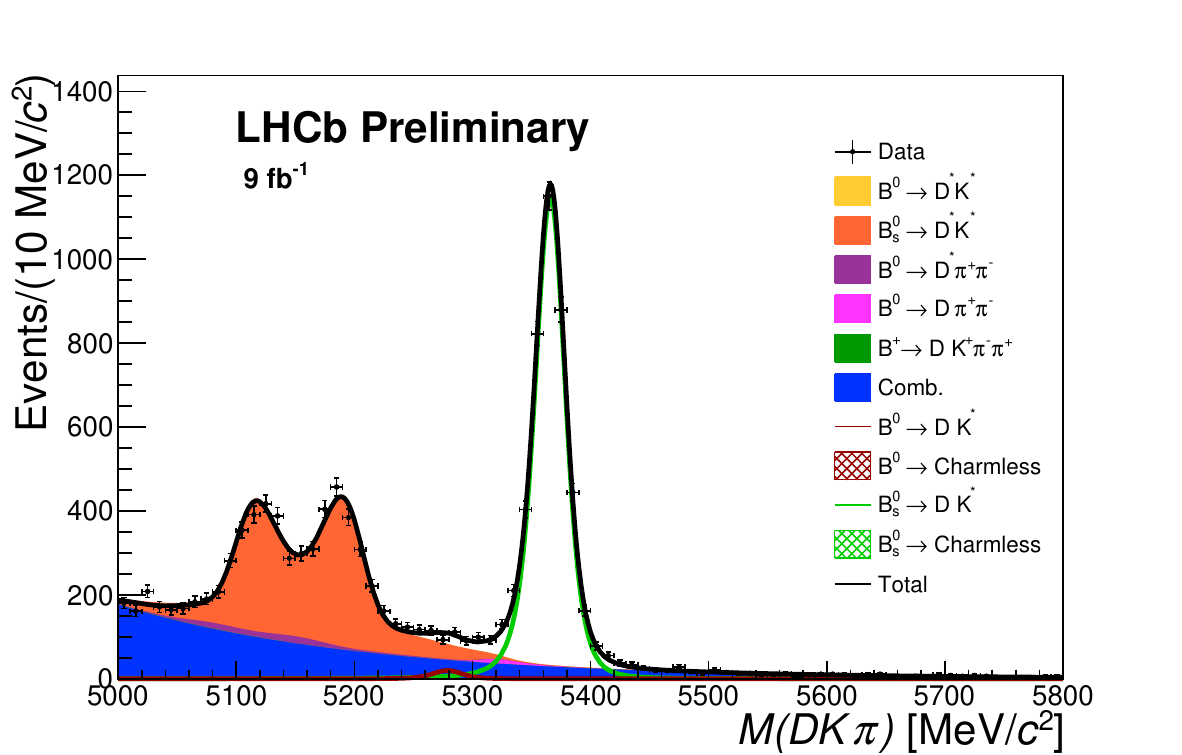}\hfill
    \includegraphics[width=0.43\textwidth]{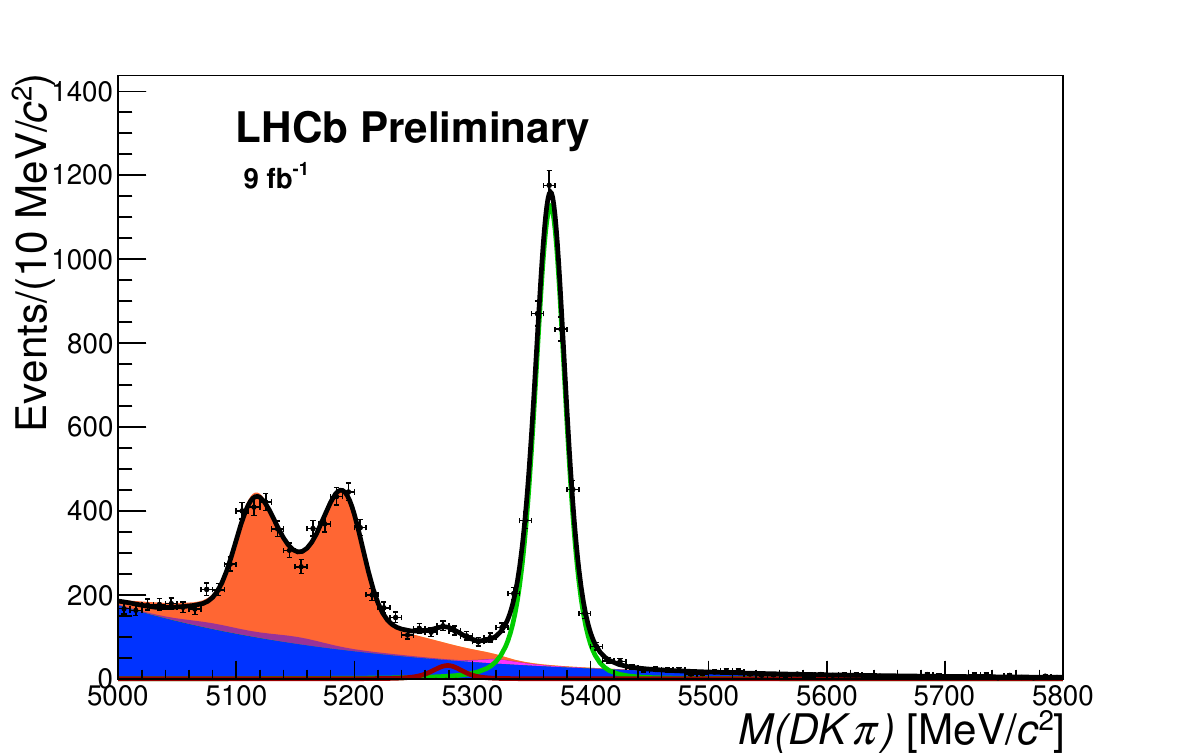} 
    \caption{Results (from top to bottom) from the $D \to K^{\pm}\pi^{\mp}$ control mode, $D \to K^{\pm}K^{\mp}$ GLW mode, $D \to \pi^{\pm}K^{\mp}$ ADS mode, $D \to \pi\pi\pi\pi$ GLW mode, and $D \to \pi K \pi\pi$ ADS mode for (left) $\overline{B^{0}}$ decays and (right) $B^{0}$ decay.}
        \label{fig:D0Kst_results}
\end{figure}

One way to measure \CP-violation is using $D$-decay final states which are \CP-eigenstates. These are studied using the GLW~\cite{GRONAU1991172,GRONAU1991483} method. In addition, quasi-GLW decays refer to final states which have a significant \CP content, meaning the amplitude for the $D^{0}$ and $\overline{D^{0}}$ decays are roughly, but not exactly, equal. This is quantified by the hadronic parameter, $r_{D}$, the ratio of the two amplitudes, and the strong-phase difference between them, $\delta_{D}$.
The amplitude of the ${B^{0}\to D^{0}K^{*0}}$ decay is smaller than that of the ${B^{0}\to \overline{D^{0}}K^{*0}}$ decay, and this is quantified by the hadronic parameter, $r_{B}$. The strong and weak phase differences between the amplitudes are given by $\delta_{B}$ and $\gamma$, respectively. This dependence on $\gamma$ indicates that interference is required to measure it. The sensitivity to $\gamma$ is proportional to the size of $r_{B}$, which for ${B^{0}\to D K^{*0}(892)}$ decays is around 0.25~\cite{LHCb-CONF-2022-003}. To measure $\gamma$, the ratio of the rate compared to a control mode, which are Cabibbo-favoured modes, and the asymmetry between the $B^{0}$ and ${\overline{B^{0}}}$ decays are measured. These are defined as follows and related to the physical parameters ($r_{B}$, $\delta_{B}$, and $\gamma$) as

\begin{equation*}
    R_{CP} \equiv \frac{\Gamma(B^{0} \rightarrow [h^{+}h^{-}]_{D}K^{*0}) + \Gamma(\overline{B^{0}} \rightarrow [h^{+}h^{-}]_{D}\overline{K^{*0}})}{\Gamma(B^{0} \rightarrow [K^{-}\pi^{+}]_{D}K^{*0}) + \Gamma(\overline{B^{0}} \rightarrow [K^{+}\pi^{-}]_{D}\overline{K^{*0}})} \frac{\mathcal{B}(D^{0} \rightarrow K^{-}\pi^{+})}{\mathcal{B}(D^{0} \rightarrow h^{+}h^{-})}
\end{equation*}
\begin{equation}
    = \frac{1 + r_{B}^{2} + 2 r_{B}\kappa\cos(\delta_{B})\cos(\gamma)}{1 + r_{B}^{2}r_{D}^{2} + 2 r_{B}r_{D}\kappa\cos(\delta_{B} - \delta_{D})\cos(\gamma)},
\end{equation}
\begin{equation}
    A_{CP} \equiv \frac{\Gamma(B^{0} \rightarrow [h^{+}h^{-}]_{D}K^{*0}) - \Gamma(\overline{B^{0}} \rightarrow [h^{+}h^{-}]_{D}\overline{K^{*0}})}{\Gamma(B^{0} \rightarrow [h^{+}h^{-}]_{D}K^{*0}) + \Gamma(\overline{B^{0}} \rightarrow [h^{+}h^{-}]_{D}\overline{K^{*0}})}
    = \frac{2 r_{B}\kappa \sin(\delta_{B})\sin(\gamma)}{1 + r_{B}^{2} + 2 \kappa\cos(\delta_{B})\cos(\gamma)}.
\end{equation}
The ratios and asymmetries are determined by fitting the mass distribution of the $D K \pi$ system from either $B^{0}$ or $\overline{B^{0}}$ decays. Results of these fits for the control modes are shown in Fig.~\ref{fig:D0Kst_results}. The asymmetry, $A_{K\pi} = 0.033 \pm 0.017 \pm 0.015$, is consistent with no asymmetry as expected. This is since the interference in this control mode is smaller than the experimental sensitivity.

For these asymmetries, there are contributions from not only interference but also production and detection asymmetries, for which corrections must be applied. The value for the production asymmetry is taken from Ref.~\cite{Aaij_2017}, $A_{\text{prod}} = (-8 \pm 5) \times 10^{-3}$. The detection asymmetry is determined using the difference in detection asymmetries for kaons and pions. This is estimated using calibration data for $D^{+}\to K^{-}\pi^{+}\pi^{+}$ and $D^{+} \to K_{S}^{0}\pi^{+}$ decays. The asymmetries are then weighted to match the signal kinematic distributions resulting in $A_{\text{det}} = (-9.8 \pm 5.5) \times 10^{-3}$. The detection asymmetry correction is then applied depending on the difference between the number of kaons and pions of the same sign in the final state~\cite{Gilman:2812327}.

There are also efficiency corrections to consider. Since the \CP observables are calculated from ratios, most efficiencies cancel but two corrections must be applied. Firstly, due to the selection. This is determined by applying the selection to simulation. Secondly, charge-dependent PID efficiencies. These are estimated using calibration data for kaons and pions from $D^{*}$ decays. These vary with track momentum, and therefore efficiencies determined using calibration data are then weighted to match the kinematics of the signal decays using simulation.

An example of the results for GLW-like decays is shown in Fig.~\ref{fig:D0Kst_results} for $D \to K^{+}K^{-}$ decays. 
Here, the asymmetry is small, $A_{\CP}^{KK} = -0.047 \pm 0.063 \pm 0.015$, but it is the rate measurement which is important, $R_{\CP}^{KK} = 0.817 \pm 0.057 \pm 0.017$. This would be close to 1 if the interference effect were small. But the rate deviates from 1 and therefore contributes significant information towards $\gamma$. Alongside the $D \to K^{+}K^{-}$ mode, the $D \to \pi^{+}\pi^{-}$ mode is also measured and results are in Ref.~\cite{Gilman:2812327}.

\subsection*{ADS decays}

The measurement in Ref.~\cite{Gilman:2812327} also studies ADS-like decays. These are final states which, depending on the $D$ flavour, occur via favoured or doubly-Cabibbo suppressed decays.  
This means that the two interfering amplitudes (the total amplitudes including the $B$ and $D$ decays) are of similar magnitudes, resulting in maximal interference and greater sensitivity to $\gamma$. These decays are measured using the ADS~\cite{PhysRevLett.78.3257,PhysRevD.63.036005} method where the \CP observables are ratios of the rate of the suppressed to the favoured modes. These are defined as,
\begin{equation}
    R^{+}_{\pi K} \equiv \frac{\Gamma(\overline{B^{0}} \rightarrow [\pi^{+}K^{-}]_{D}\overline{K^{*0}})}{\Gamma(\overline{B^{0}} \rightarrow [K^{+}\pi^{-}]_{D}\overline{K^{*0}})} 
    = \frac{r_{B}^{2} + r_{D}^{2} + 2 r_{B}r_{D}\kappa\cos(\delta_{B} + \delta_{D} + \gamma)}{ 1 + r_{B}^{2} r_{D}^{2} + 2 r_{B}r_{D}\kappa\cos(\delta_{B} - \delta_{D} + \gamma)},
\end{equation}
\begin{equation}
    R^{-}_{\pi K} \equiv \frac{\Gamma(B^{0} \rightarrow [\pi^{-}K^{+}]_{D}K^{*0})}{\Gamma(B^{0} \rightarrow [K^{-}\pi^{+}]_{D}K^{*0})} 
    = \frac{r_{B}^{2} + r_{D}^{2} + 2 r_{B}r_{D}\kappa\cos(\delta_{B} + \delta_{D} - \gamma)}{ 1 + r_{B}^{2} r_{D}^{2} + 2 r_{B}r_{D}\kappa\cos(\delta_{B} - \delta_{D} - \gamma)}.
\end{equation}

The results from this mode with $D \to \pi^{\pm} K^{\mp}$ decays are shown in Fig.~\ref{fig:D0Kst_results} where $R^{+}_{\pi K} = 0.069 \pm 0.013 \pm 0.005$ and $R^{-}_{\pi K} = 0.093 \pm 0.013 \pm 0.005$. To leading order, $R^{\pm} \approx r_{B}^{2}$. Therefore, larger rates imply a larger $r_{B}$.

\subsection*{Multi-body $D$ decays}

In this measurement~\cite{Gilman:2812327}, multi-body $D$-decay final states are also studied. The GLW-like decay is $D \to \pi^{+}\pi^{-}\pi^{+}\pi^{-}$, where the \CP-even fraction is $F_{+} = 0.735 \pm 0.015 \pm 0.005$~\cite{PhysRevD.106.092004}. This alters the \CP observables by a factor of $(2F_{+}-1)$, for example, 
\begin{equation}
    A_{CP} =  \frac{2 r_{B}\kappa \textcolor{red}{(2 F_{+} - 1)}\sin(\delta_{B})\sin(\gamma)}{1 + r_{B}^{2} + 2 r_{B}\kappa\cos(\delta_{B})\cos(\gamma)}.
\end{equation}

For the ADS mode, $D \to \pi^- K^+ \pi^- \pi^+$ decays are used and this requires an additional coherence factor, $R_{K3\pi}$, due to intermediate resonances. The value for this is taken from  Ref.~\cite{2021}, and is $R_{K3\pi} = 0.44 ^{+0.10}_{-0.09}$. This alters the decay rates, for example,
\begin{equation}
        \Gamma(B^{0} \rightarrow [K^{+}\pi^{-}\pi^{+}\pi^{-}]_{D}K^{*0}) \propto r_{B}^{2} + r_{D}^{2} + 2 r_{B}r_{D}\kappa  \textcolor{red}{R_{K3\pi}}\cos(\delta_{B} + \delta_{D} - \gamma),
\end{equation}
and therefore reduces the interference by a factor of 2.

The results from the GLW mode are shown in Fig.~\ref{fig:D0Kst_results}, where again a small asymmetry is seen, $A_{\CP}^{4\pi} = 0.014 \pm 0.087 \pm 0.016$, and the rate is $R_{\CP}^{4\pi} = 0.882 \pm 0.086 \pm 0.033$. The results from the ADS mode are shown in Fig.~\ref{fig:D0Kst_results}, where $R^{+}_{\pi K\pi\pi} = 0.060 \pm 0.014 \pm 0.005$ and $R^{-}_{\pi K \pi \pi} = 0.038 \pm 0.014 \pm 0.005$.

\subsection*{Results}

In general the statistical uncertainties on the \CP observables from Ref.~\cite{Gilman:2812327} have improved by 60\% compared to the previous analysis~\cite{LHCb-PAPER-2019-021} due to the increased signal yield. For both the two- and four-body $D$-decays, the dominant source of systematic uncertainty in the GLW asymmetries is the asymmetry correction, and in the GLW ratios it is due to external measurements of the $D^{0}$ branching fractions. In the ADS mode there is no dominant source of systematic uncertainty. Overall, the analysis is statistically limited.

\begin{figure}[h]
        \centering
        \includegraphics[width=0.45\textwidth]{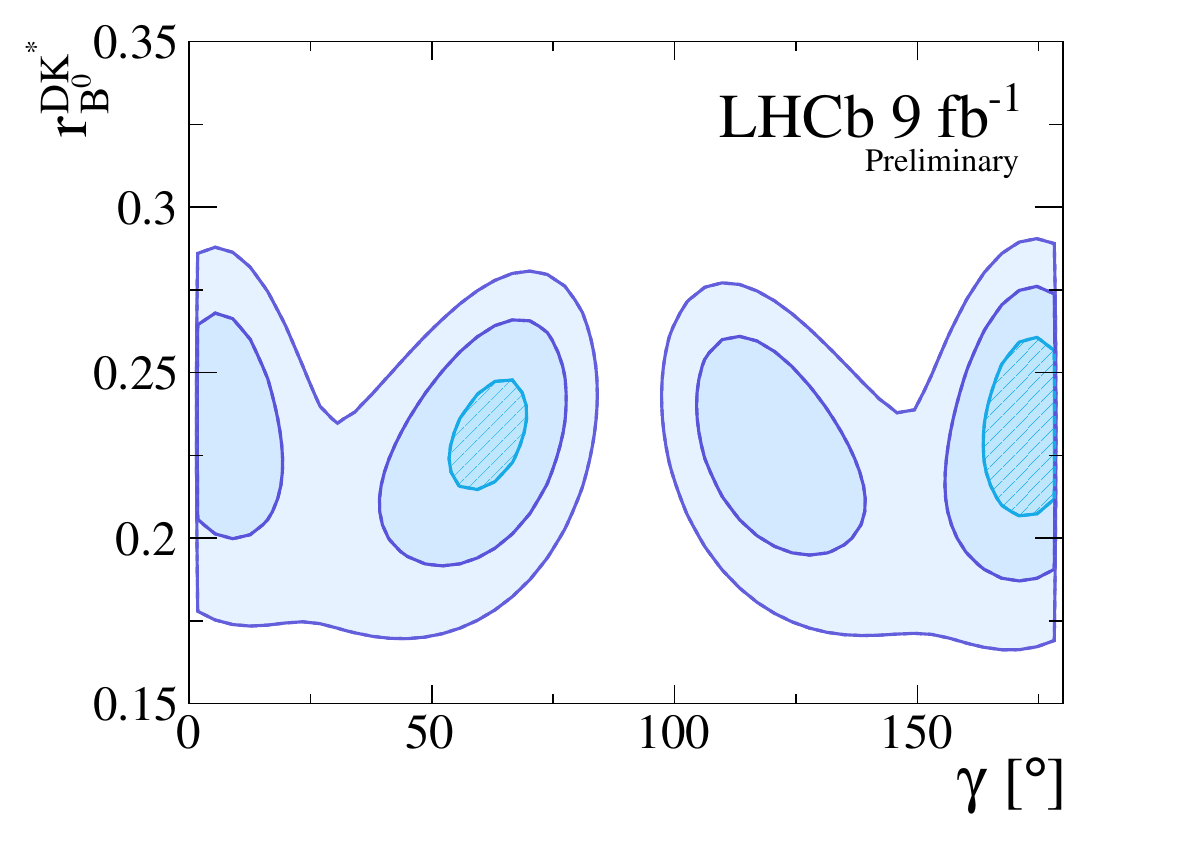}
        \includegraphics[width=0.45\textwidth]{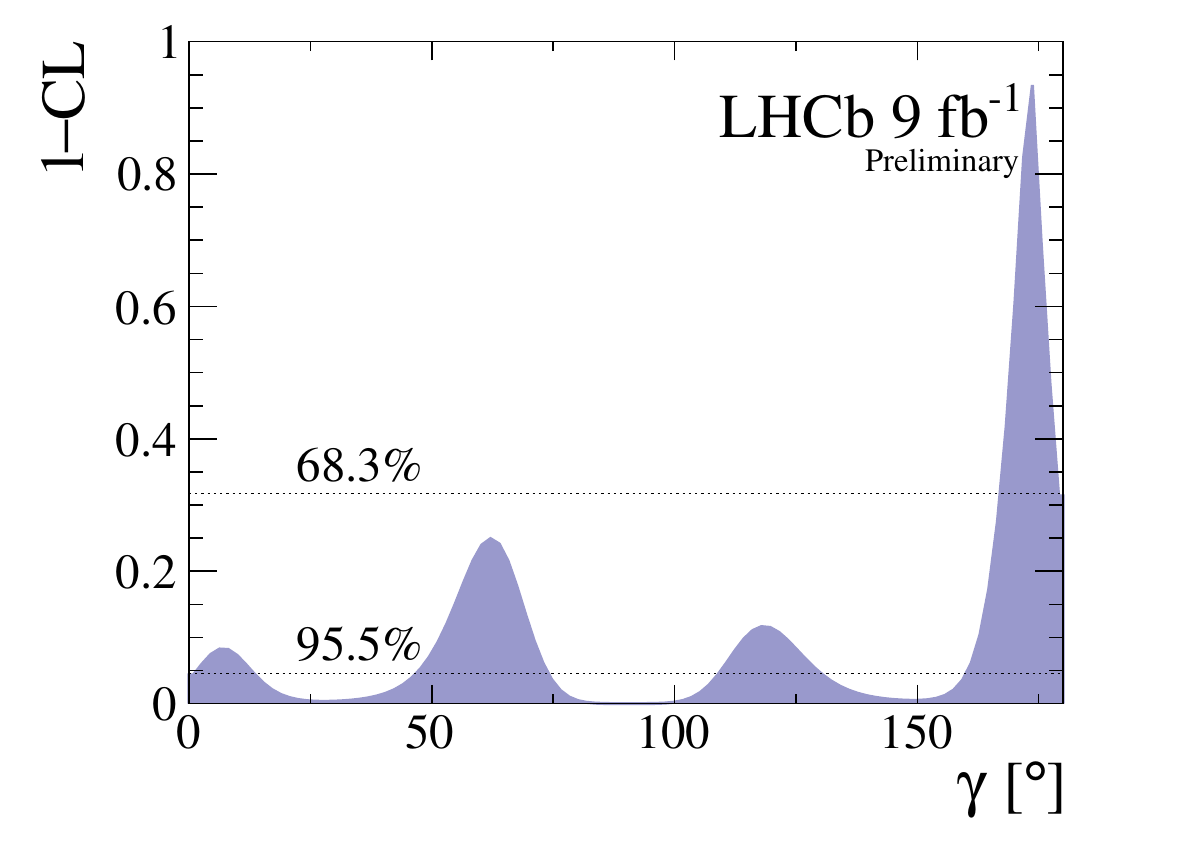}
        \caption{The 68.3\% and 95.5\% confidence levels from Ref.~\cite{Gilman:2812327} in (left) $r_{B^0}^{DK^{*}}$ and $\gamma$, and (right) $\gamma$.}
        \label{fig:D0Kst_gamma}
    \end{figure}

The \CP observables can be interpreted in terms of the physical observables using the procedure detailed in Ref.~\cite{LHCb-PAPER-2016-032}. The trigonometric functions in the relations of the \CP observables and the low sensitivity to asymmetries mean that there is a four-fold degeneracy, as shown in Fig.~\ref{fig:D0Kst_gamma}. The preferred solution of $\gamma$, as shown on the right of Fig.~\ref{fig:D0Kst_gamma}, is at $172 \pm 6 \degrees$ which includes both statistical and systematic uncertainties. However, an alternative solution, which is consistent with the world average of direct measurements, is at $62 \pm 8 \degrees$. Results from the $B^{0} \to D K^{*0}$ analysis with $D \to K_{s}^{0} \pi\pi$ and $D \to K_{s}^{0} KK$ final states~\cite{lhcbcollaboration2023measurement} are needed to break this degeneracy.

\section*{$B^{\pm}\to D K^{\pm}$ and $B^{\pm}\to D \pi^{\pm}$ decays with $D \to K\pi\pi\pi$}

A measurement of \CP-violation in $B^{\pm}\to D K^{\pm}$ and $B^{\pm}\to D \pi^{\pm}$ decays with $D \to K\pi\pi\pi$ final states was performed by \lhcb~\cite{2023}. The larger yields in $B^{\pm} \to D K^{\pm}$ compared to $B^{0} \to D K^{*0}$ mean that the sensitivity to $\gamma$ can be improved by splitting the $D$-decay phase space into bins where the coherence factor is larger. This was proposed in Ref.~\cite{EVANS} and the binning determined in Ref.~\cite{2021} is used.

\subsection*{Results}
\begin{figure}[h!]
\centering
    \includegraphics[width=0.49\textwidth,trim={0 0.4cm 0cm 1cm},clip]{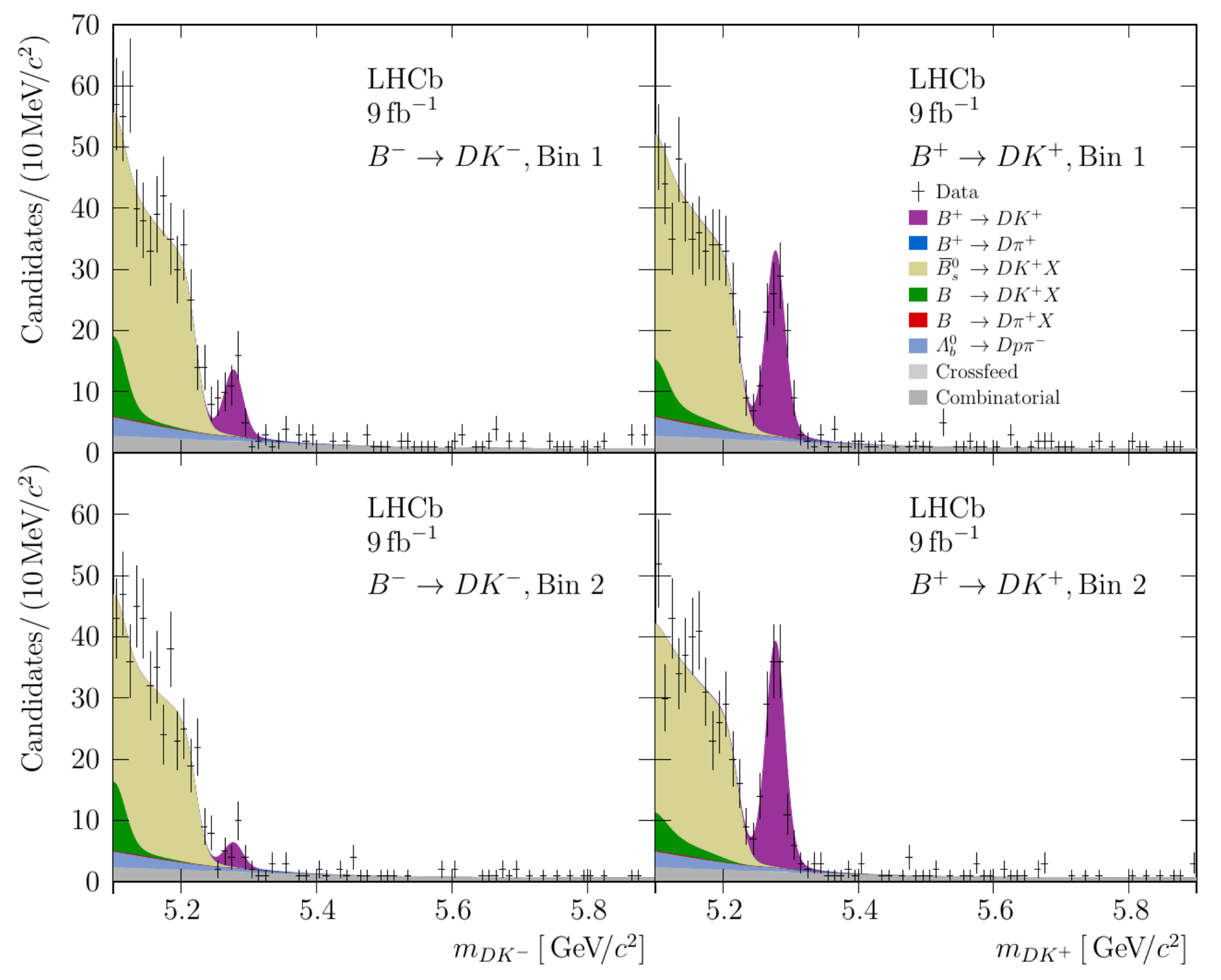}
    \includegraphics[width=0.5\textwidth,trim={0 0 0cm 0.07cm},clip]{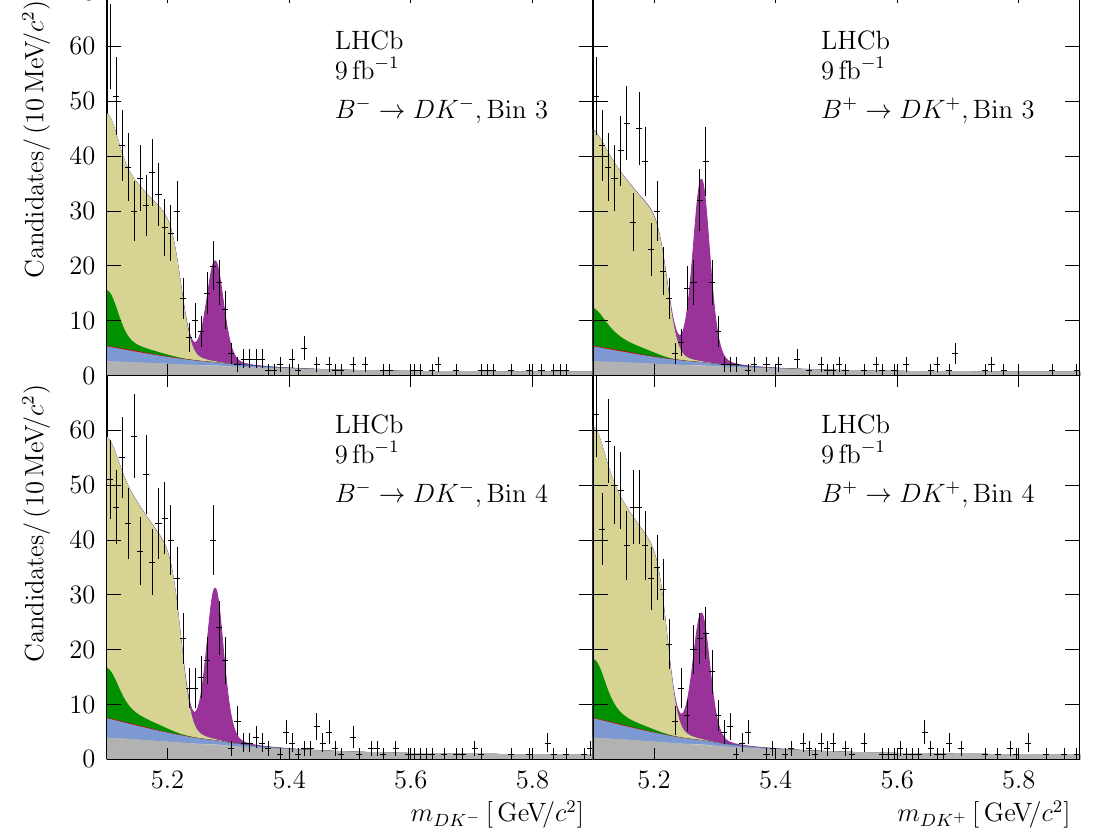}
        \caption{Binned mass distributions in $m(DK^-)$ and $m(DK^+)$ from Ref.~\cite{2023}.}
        \label{fig:K3pi_results}
\end{figure}

The asymmetries are determined using fits to mass spectra of the $DK^{+}$ or $DK^{-}$ system. Results for this in each bin are shown in Fig.~\ref{fig:K3pi_results}, where bin 2 shows the largest \CP asymmetry seen at \lhcb. The bins labelled in these plots are ordered by the average value of $\delta_{D}$. This means that from bin 1 to 4, the asymmetry can be seen to rise and fall.

Using these measurements, $\gamma$ is found to be, $\gamma = (54.8 ^{+6.0}_{-5.8}$$^{+0.6}_{-0.6}$$^{+6.7}_{-4.3})\degrees$, where the first uncertainty is statistical, the second is systematic and the third is due to external inputs. The last uncertainty is slightly larger than the statistical uncertainty and updated external inputs are needed to improve this.

\section*{Conclusion}
Two recent LHCb measurements of the CKM angle $\gamma$ which study ADS and GLW-like decays have been presented. These measurements along with other recent analyses from LHCb show that the target precision on $\gamma$ of $4\degrees$ has been surpassed. This will continue to improve as on-going analyses are completed and Run 3 data is analysed.



\end{document}